\documentclass[pre,twocolumn,showpacs,preprintnumbers,amsmath,amssymb,superscriptaddress]{revtex4}

\usepackage{graphicx}
\usepackage{dcolumn}
\usepackage{epsfig}

\newcommand{\be}{\begin{equation}}
\newcommand{\ee}{\end{equation}}
\newcommand{\bd}{\begin{displaymath}}
\newcommand{\ed}{\end{displaymath}}
\newcommand{\BE}{\begin{eqnarray}}
\newcommand{\EE}{\end{eqnarray}}

\newcommand{\erf}{{\rm erf}}
\newcommand{\erfc}{{\rm erfc}}

\newcommand{\ba}{\ensuremath{\mathbf{a}}}

\newcommand{\avg}[1]{\left\langle{#1}\right\rangle}

\begin{document}


\title{Effects of Tobin Taxes in Minority Game markets}
\author{Ginestra Bianconi}
\email{gbiancon@ictp.trieste.it}
\affiliation{The Abdus Salam International Centre for Theoretical Physics, Strada Costiera 11, 34014 Trieste, Italy}
\author{Tobias Galla}
\email{galla@ictp.trieste.it}
\affiliation{The Abdus Salam International Centre for Theoretical Physics, Strada Costiera 11, 34014 Trieste, Italy}
\affiliation {CNR-INFM, Trieste-SISSA Unit, V. Beirut 2-4, 34014 Trieste, Italy
}
\author{Matteo Marsili}
\email{marsili@ictp.trieste.it}
\affiliation{The Abdus Salam International Centre for Theoretical Physics, Strada Costiera 11, 34014 Trieste, Italy}

\date{\today}

\begin{abstract}
We show that the introduction of Tobin taxes in agent-based models of
currency markets can lead to a reduction of speculative trading and
reduce the magnitude of exchange rate fluctuations at intermediate tax
rates. In this regime revenues for the market maker obtained from
speculators are maximal. We here focus on Minority Game models of
markets, which are accessible by exact techniques from statistical
mechanics. Results are supported by computer simulations. Our findings
suggest that at finite systems sizes the effect is most pronounced in
a critical region around the phase transition of the infinite system,
but much weaker if the market is operating far from criticality and
does not exhibit anomalous fluctuations.
\end{abstract}
\pacs{PACS}

\maketitle

\section{Introduction}

In 1972 James Tobin proposed to ``throw some sand in the wheels of
our excessively efficient international money markets''
\cite{tobin78} by imposing a tax of 0.05 to 0.5\% on all
international currency transactions. The Bretton Woods agreement -- a
system of fixed foreign exchange rates tied to the price of gold -- at
that time was gradually being dismantled, with the USA stepping out in
1971. This system had been introduced in the wake of World War II in
order to rebuild global capitalism. Tobin feared the effects of
countries exposed to freely fluctuating exchange rates and suggested,
as a second best solution, the introduction of what is now called
a `Tobin tax' in order to suppress speculative behaviour and thus
containing exchange rates volatility within tenable levels.

Since then, under floating exchange rates, the trading volume on
international currency markets has grown sharply, especially after
the introduction of electronic trading, reaching a level of $1.5$
trillion US-Dollar per day in 2000 \cite{ehrenstein}. Most of
these transactions are effected on time-scales of less than seven
days, more than $40\%$ involve round-trips within two days or
less, and $90\%$ are of a speculative nature, only one tenth are
carried out within the production sector \cite{ehrenstein}.

While the Tobin tax has never been implemented in reality the
discussion of this issue is still lively and opinions are widely
divided between proponents who claim that a Tobin tax would improve
the situation of countries damaged by international currency
speculation \cite{ramonet}, and opponents who reject the proposal on
various grounds. They claim that its implementation would hardly be
feasible and extremely expensive, that it would mostly damage
developing countries and that through a reduction of market liquidity,
Tobin taxes might indeed result in more, not less volatility
\cite{FBELetter}.

The question of whether a tax on financial transactions reduces
volatility or not is indeed a subtle one. On the one hand excess
volatility is related to trading volume, hence reducing the currency
trading activity through the introduction of a tax on currency
transactions can be expected to reduce volatility.  On the other hand
speculators provide liquidity and eliminate market inefficiencies,
hence volatility might increase when a tax is levied and speculators
drop out of the market due to the increasing trading costs. The latter
consideration is particularly relevant given that the margin on which
speculators live is extremely small and even a $0.5\%$ tax could turn
their marginal gains negative.

In order to make these considerations quantitative some authors have
investigated the effects of Tobin taxes in agent-based models of
financial markets \cite{ehrenstein,ehrenstein2,westerhoff}. These
model are capable to reproduce typical features of price fluctuations
in real markets -- so-called stylized facts \cite{stylized_facts} --
and are therefore well suited to shed light on the effects of
introducing a Tobin tax. In the context of zero-intelligence
percolation models
\cite{contbouchaud}, Ehrenstein and co-workers found that
generally the introduction of Tobin tax brings about a reduction
in volatility, as long as the tax rate is not too high to cause
liquidity problems \cite{ehrenstein, ehrenstein2}. Westerhoff
\cite{westerhoff} comes to similar conclusions building on the
Chiarella-Hommes approach of modeling financial markets with
systems of heterogeneous agents \cite{CH}. At variance with
previous models, this study introduces considerations of agents'
rationality through heuristic expectation models of future returns
of a chartist or fundamentalist nature.

The present paper addresses similar general questions. In contrast
with previous works, however, we take the view of a financial market
as an ecology of different types of agents interacting along an
`information food chain'. In our picture speculators `predate' on
market inefficiencies (arbitrage opportunities) created by other
investors (the so-called producers). This approach has been recently
formalized in the Minority Game (MG) \cite{Book1,Book2,Book3}. The MG
captures the interplay between volatility and market efficiency in a
vivid though admittedly simplified and stylized way. Indeed the
analysis of the MG has revealed that within this model framework
excess volatility and market efficiencies are identified as two sides
of the same coin, both resulting as consequences of speculative
trading. The MG exhibits two different regimes, one in which the
market is fully efficient and another in which arbitrage opportunities
are not entirely eliminated by the dynamics of the agents. These
regimes are separated sharply in the parameter space of the model, and
it turns out that the boundary at which the market becomes efficient
coincides precisely with the locus of a phase transition in the
language of statistical physics. At the same time critical
fluctuations -- very similar to the stylized facts observed in real
market data -- emerge in the vicinity of this transition but not
further away \cite{GCMG}. Hence, at odds with previous models, the MG
might offer a perspective of understanding how the introduction of
transactions taxes affects the information ecology and market
efficiency. It will here be important to distinguish between regimes
close to and far away from the above transition between efficient and
inefficient phases of the market. A further advantage of the MG over
other more elaborate agent-based models lies in its analytic
tractability. Despite its stylized setup the phenomenology of the MG
is remarkably rich, but at at the same time the MG model can be
understood fully analytically with tools of statistical physics of
disordered systems
\cite{Book1,Book2}. This analytical solution provides an understanding
of the model, which goes much deeper than approaches to other models
purely based on numerical simulations. In particular it is possible to
derive closed analytical expressions for key observables such as the
market volatility, the trading activity of the agents and the revenue
for the market maker.

In brief, our main result is that within the picture of the MG model a
small tax decreases volatility whenever a {\em finite} market exhibits
anomalous fluctuations. Indeed the fundamental effect of the tax is to
draw the market away from its critical point. At the same time, the
tax introduces an information inefficiency and thus too high a tax
might not be advisable. The total revenue from the tax exhibits a
maximum for intermediate rates similar to what was found in earlier
works on different models \cite{ehrenstein}. Furthermore, the effects
of imposing a tax materialize in the market behaviour only after times
which scale inversely with the tax rate.  Extremely small tax rates
may thus need a long time to stabilize turbulent markets.  This can be
quite relevant if this time scale becomes comparable to that over
which market's composition changes. In particular, if speculators
leave and enter the population of traders at a too fast rate a small
Tobin tax may fail to stabilize the market. Large taxes in such a
scenario reduce the time the population needs to co-ordinate below the
rate with which new agents enter the market. Under such circumstances
a transaction tax may thus prevent high volatility states and reduce
the volatility significantly even in an {\em infinite} system.

In the following we shall first introduce the grand-canonical MG
(GCMG) and re-iterate its known main features. In the main sections we
then comment on how a tax on transactions can be introduced and
discuss the effects on the market within the present model. We then
turn to a brief discussion on how to relate these results to real
markets, summarize our results and give some final concluding remarks.

\section{The Grand-canonical Minority Game}

\subsection{The model}

The so-called grand-canonical MG (GCMG) describes a simple market of
$N$ agents $i=1,\dots,N$ who at each round of the game make a binary
trading decision (to buy or to sell) or who each may decide to refrain
from trading. They thus each submit bids $b_i(t)\in\{-1,0,1\}$ in
every trading period $t=0,1,2,\dots$ resulting in a total excess
demand of $A(t)=\sum_{i=1}^N b_i(t)$.

These trading decisions are taken to be based on a stream of
information available to the agents. This common information on the
state of the market (or on other questions relevant to the market) is
encoded in an integer variable $\mu(t)$ at time $t$ taking values in
$\{1,2,\ldots,P\}$ \cite{nota1}. Here we assume
that $\mu(t)$ models an exogenous news arrival process, and that the
$\{\mu(t)\}$ are drawn at random from the set $\{1,\ldots,P\}$,
independently and with equal probabilities at each time \cite{nota2}.  The objective of each agent is to be in the minority
at each time-step, i.e. to place a bid $b_i(t)$ which has the opposite
sign of the total bid $A(t)$. This minority setup corresponds to
contrarian behaviour and can be derived from a market mechanism taking
into account the expectations of the traders on the future behaviour
of the market \cite{M01}. In order to do so, each agent has a `trading
strategy' at his disposal. Trader $i$'s strategy is labelled by
$\ba_i=(a_i^\mu)_{\mu=1,\dots,P}\in\{-1,1\}^P$ and provides a map from
all values of the information $\mu$ onto the binary set $\{-1,1\}$ of
actions (buy/sell). Upon receiving information $\mu$ the trading
strategy of agent $i$ thus prescribes to take the trading action
$a_i^\mu\in\{-1,1\}$.  These strategies are assigned at random and
with no correlations at the beginning of the game, and then remain
fixed \cite{nota3}. Agents in the GCMG are adaptive and may decide not to trade if
they do not consider their strategy adequate. More precisely, each
agent keeps a score $u_i(t)$ measuring the performance of his strategy
vector. He then trades at a given time-step $t$ only if his strategy
has a positive score $u_i(t)>0$ at that time. Therefore, the bids of
agents take the form $b_i(t)=n_i(t)a_i^{\mu(t)}$ with $n_i(t)=1$ if
$u_i(t)>0$ and $n_i(t)=0$ otherwise. Accordingly the excess demand is
given by

\be
A(t)=\sum_{i=1}^N n_i(t)a_i^{\mu(t)}. \label{At}
\ee

Agent $i$ keeps a record of the past performance of his strategy $a_i^\mu$ by
updating the score $u_i(t)$ as follows

\be \label{eq:update}
u_i(t+1)=u_i(t)-a_i^{\mu(t)}A(t)-\varepsilon_i.
\ee

\noindent at each step, with constant $\varepsilon_i$. The first
term $-a_i^{\mu(t)}A(t)$ is the Minority Game payoff, it is positive
whenever the trading action $a_i^{\mu(t)}$ proposed by $i$'s strategy
vector and the aggregate bid $A(t)$ are of opposite signs, and
negative whenever $i$ joins the majority decision. The idea of
Eq. (\ref{eq:update}) is that whenever the payoff $-a_i^{\mu(t)}A(t)$
is larger than $\varepsilon_i$ the score of player $i$'s strategy is
increased, otherwise it is decreased. The constant $\varepsilon_i$ in
(\ref{eq:update}) thus captures the inclination of agent $i$ to trade
in the market. This inclination will in general be heterogeneous
across the population of agents, with agents with high values of
$\varepsilon_i$ being more cautions to trade than agents with low
$\varepsilon_i$. In our simplified model we only consider two types of
agents. First we assume that there are $N_s\le N$ speculators who
trade only if their perceived market profit obtained by using their
strategy exceeds a given threshold, and hence we set
$\varepsilon_i=\epsilon\ge 0$ for such agents. Here $\epsilon$ can be
considered as the speculative margin of gain in a single transaction.

The remaining $N_p=N-N_s$ agents -- the so-called producers or
institutional investors -- are assumed to trade no matter
what. Mathematically this is implemented by setting
$\varepsilon_i=-\infty$ for this second group of agents. They have
$n_i(t)=1$ for all times $t$. For convenience we will order the agents
such that speculators carry the indices $i=1,\dots,N_s$ and producers
the labels $i=N_s+1,\dots, N$. 

\subsection{Price process, volatility and predictability} 

Within the MG setup the market volatility is then given by

\begin{equation}\label{sigma}
  \sigma^2=\frac{\avg{A(t)^2}}{N},
\end{equation}

\noindent where $\avg{\ldots}$ will stand for a time-average in
the stationary state of the model from now on. The normalization to
the number of agents $N$ is here introduced to guarantee a finite
value of $\sigma^2$ in the infinite-system limit, with which the
statistical mechanics analysis of the model is concerned.

The information variable $\mu(t)$ allows one to quantify
information-efficiency of the model market by computing the
predictability

\begin{equation}\label{H}
  H=\frac{1}{PN}\sum_{\mu=1}^P\avg{A|\mu}^2
\end{equation}
where $\avg{\ldots|\mu}$ denotes an average conditional on the
occurrence of information pattern $\mu(t)=\mu$. A value $H\not =
0$ indicates that for some $\mu$ the minority payoff is
statistically predictable $\avg{A(t)|\mu}\not =0$, whereas the
market is unpredictable and fully efficient when $H=0$.

The simplest way to relate this picture to a financial market is to
postulate a simple linear impact of $A(t)$ on the (logarithmic) price
(or the exchange rate), i.e to assume that

\begin{equation}\label{pt}
  p(t+1)=p(t)+\frac{A(t)}{\lambda},
\end{equation}

\noindent where $\{p(t)\}$ denotes a price (exchange rate) process and where $\lambda$ 
is the liquidity. In a derivation of Eq. (\ref{pt}) from a market
clearing condition $\lambda$ turns out to be inversely proportional to
the number of active traders $\sum_i n_i(t)$ \cite{M01}. In the
following we set the liquidity to $\lambda=\sqrt{N}$, so that
$\sigma^2$ becomes the volatility of the price process,
$\sigma^2=\avg{(p(t+1)-p(t))^2}$. It is found simulations that the
effects of introducing a Tobin tax on the behaviour of the model are
qualitatively similar for either choice of $\lambda$, so that we will
stick with the technically more convenient first definition.

\subsection{The behaviour of the GCMG}

\begin{figure}[t]
\hspace*{35mm} \setlength{\unitlength}{1.1mm}
\begin{picture}(100,35)
\put(-26,0){\epsfysize=40\unitlength\epsfbox{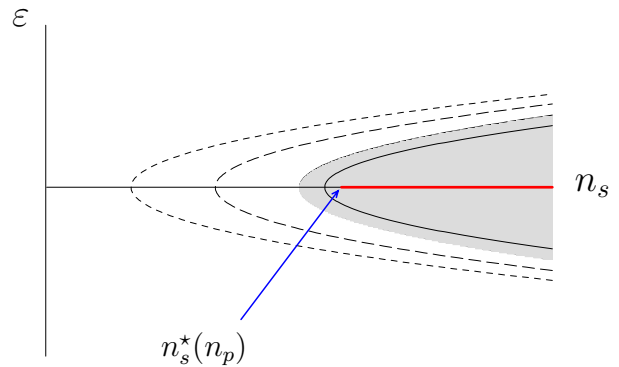}}
\put(38,20){\Large $n_s$}
\put(-30,40){\Large $\varepsilon$}
\put(-12,0){\large $n_s^\star(n_p)$}
\end{picture} 
\caption{Phase diagram of the GCMG in the $(n_s,\varepsilon)$-plane at fixed $n_p$. The red line segment at $\varepsilon=0$ and $n_s\geq n_s^\star(n_p)$ marks the phase transition in the limit of infinite system size. At finite size anomalous fluctuations and stylized facts are found in a region around this critical line, as indicated by the shaded area. This so-called `critical region' indicates regions with strong dynamical and finite size effects and is large for small systems and shrinks towards the critical line segment in the infinite-size limit. }
\label{fig:phasediagram}
\end{figure}

The GCMG has been studied in great detail in \cite{GCMG,finitememory} with methods
well established in statistical mechanics. The analysis is here
generally concerned with the stationary states of the system, i.e. the
behaviour which is reached after running the learning dynamics
of the agents for some sufficiently long transient equilibration time.

The statistical mechanics approach provides exact results for the
model in the limit of infinite market sizes, where one takes the
number of agents $N=N_s+N_p$ and the number $P$ of possible different
information states to infinity, while at the same time keeping the
ratios $n_s=N_s/P$ and $n_p=N_p/P$ fixed and finite.  $n_s$ and $n_p$
along with $\varepsilon$ are thus control parameters of the
model. This approach \cite{GCMG,Book1,Book2} makes it possible to
derive exact expressions for several quantities, including the
predictability $H$ and upon neglecting time-dependent correlations
accurate approximations for the volatility $\sigma^2$ can be found
\cite{MC01,HeimCool01}. We will here not enter the detailed
mathematics of the calculations, the resulting equations for the key
quantities in the stationary states as well as a sketch of their
derivation are found in the appendix. Further details regarding the
statistical mechanics analysis of MGs and GCMGs can be found e.g. in
\cite{GCMG,Book1,Book2} and in references therein.

The overall picture which emerges is the following \cite{GCMG}: at
fixed $n_p$, the statistical behavior of the model is characterized by
a critical line at $\epsilon=0$ which extends from some critical value
$n_s^\star(n_p)$ to larger values $n_s\ge n_s^\star(n_p)$ than this
threshold. This is illustrated in Fig. \ref{fig:phasediagram}. As this
line (segment) is approached in parameter space the market becomes
more and more efficient, i.e. $H\to 0$ as $\varepsilon\to 0$ for
$n_s\ge n_s^\star$. On the critical line $(\epsilon=0, n_s\ge
n_s^\star)$ itself the market is fully efficient and one finds $H=0$
exactly in the limit of infinite system size. In addition, numerical
simulations of the model at finite sizes close to the critical line
reveal fluctuation properties which are similar to those observed in
real markets \cite{stylized_facts}. In particular $A(t)$ has a fat
tailed distribution and one observes volatility clustering. These
effects become weaker as the system size is increased at constant
values of $n_s,n_p$ and $\varepsilon$, and similarly they disappear
gradually when one moves away from the critical line at fixed system
size.

The case $\epsilon=0$ and $n_s>n_s^\star(n_p)$ is peculiar because it
turns out that the stationary state is here not unique, but rather
that it depends on the initial conditions from which simulations are
started. In the following we will not consider the case of a strictly
vanishing $\varepsilon$, but will assume instead that speculators have
a positive profit margin $\varepsilon>0$, even if the latter may be
small. All simulations on which we report are started from zero
initial conditions $u_i(t=0)=0$ for $i=1,\dots,N_s$.
\begin{figure}[t!]
\begin{center}
\includegraphics[width=6cm]{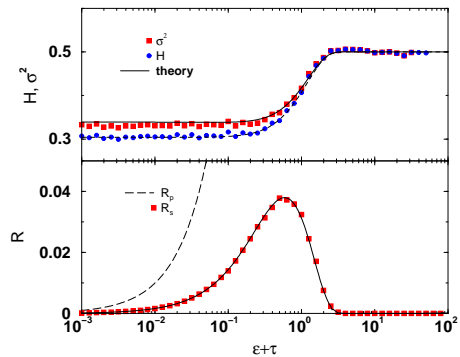}
\caption{Volatility, predictability (top) and revenue from
speculators and producers (bottom) for $n_s=n_p=1$. Symbols are data
obtained from numerical simulations with $PN_s=6000$, every data point is an average over at least $1000$
samples, simulations are run for $1000P$ steps,
with measurements taken in the second half of this interval.  Lines are the corresponding predictions
for infinite systems obtained from the analytical theory. \label{fig:ns1}}
\end{center}
\end{figure}

We finally remark that, a detailed analysis of the transient
dynamics demonstrates that for small $\epsilon$ the stationary
state is reached after a characteristic equilibration time which
scales as $1/\epsilon$ \cite{Damien_private_comm}. This long equilibration 
time is responsible for some
relevant effects if the market composition changes in time, as discussed
in Section \ref{sec:changing_agents}.

\section{Tobin tax in the GCMG}

Within the model setup the introduction of a tax $\tau$ on each
transaction can be accounted for by a change $\varepsilon_i\to
\varepsilon_i+\tau$ for all $i=1,\dots,N$. Indeed by raising $\varepsilon$ by an amount $\tau$, an additional cost $\tau$ incurs for any given agent every time they trade and no costs for agents who refrain from trading. Note that the trading volume of any fixed (active) agent is one unit in our simple setup so that $\tau$ indeed corresponds to a transaction tax per unit traded. Hence we will assume

\begin{equation}\label{eta}
\varepsilon_i=\left\{
\begin{array}{cc}
\epsilon+\tau & i\le N_s \\
-\infty & i>N_s
\end{array}\right.
\end{equation}
in the following. While this will discourage speculators from trading
(via the reduction of their strategy score) such a tax will have no
effect on the participation of producers. They will trade at every time
step as before ($n_i(t)=1$ $\forall i>N_s$ at all times $t$).

The total revenue from this tax $\tau$ is then given by

\begin{equation}\label{eq:revenue}
R=\frac{\tau}{N}\sum_i \avg{n_i(t)}=\frac{\tau}{N}\sum_{i=1}^{N_s} \avg{n_i(t)} +\tau
\frac{n_p}{n_s+n_p}\equiv R_s+R_p,
\end{equation}
where the first term corresponds to the revenue $R_s$ from
speculators and the second $(R_p)$ to that obtained from the producers.

An evaluation of the effects of levying a tax $\tau$ on the GCMG then
amounts to studying the behavior of the model as a function of
$\epsilon$ at fixed model parameters $n_s$ and $n_p$. It turns out
that here one has to distinguish between two different regimes, namely
close and far away from the phase transition. 

\begin{figure}[t!]
\begin{center}
\includegraphics[width=6cm]{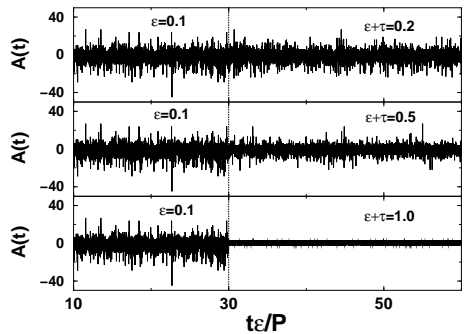}
\caption{Effect of introducing a tax on the exchange rate
fluctuations $A(t)$. Each panel corresponds to a single run with
parameters $n_s=20$, $n_p=1$, $PN_s=16000$, $\varepsilon=0.1$. In the
initial period up to $t=30P/\varepsilon$ no tax is imposed
($\tau=0$). At $t\varepsilon/P=30$ a tax rate $(\tau>0$ is introduced
and then kept fixed for the rest of the simulation. The tax rate
increases from top ($\varepsilon+\tau=0.2$) to bottom
($\varepsilon+\tau=1$).
\label{fig:timeseries}.}
\end{center}
\end{figure}

Fig. \ref{fig:ns1} reports the effects of introducing a tax on markets
whose parameters are far from the critical line. We here consider
$n_p=1$ and $n_s=1<n_s^\star(n_p=1)\approx 4.15\dots$ so that one operates
sufficiently far to the left of the critical region depicted in
Fig. \ref{fig:phasediagram}.  For such parameter values the results of
numerical simulations follow the curves predicted by the analytical
theory perfectly, and no anomalous fluctuations are present in the
corresponding price time-series. The tax has very mild effect both on
volatility and on the information efficiency, as long as $\epsilon+\tau\ll
1$. Fig. \ref{fig:ns1} also shows that the contribution of speculators
to the tax revenue has a peak at intermediate tax rates, but that at the same time the revenue $R_s$ obtained from speculators is smaller than that from institutional investors.

\begin{figure}
\begin{center}
\includegraphics[width=6cm]{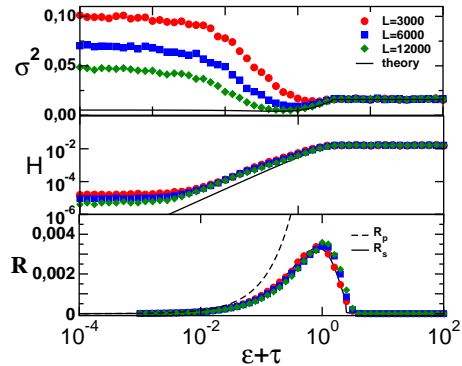}
\caption{Volatility (top), predictability (middle) and revenue (bottom) from
speculators and producers  for $n_s=60$ and $n_p=1$.  Markers are
data obtained from numerical simulations of systems with different (effective)
size $L=PN_s$  with circles, squares and diamonds corresponding to $L=3000,\,6000,\,12000$ respectively. Every data point represents an average over at least $1000$ different strategy assignments. Simulations are run for $400P+20Ns/(\varepsilon+\tau)$ steps, with measurements taken in the second half of this interval.  Lines are the predictions of the analytical theory for the infinite system. \label{fig:ns60}}
\end{center}
\end{figure}

Fig. \ref{fig:timeseries} on the other hand illustrates the response
of the market at the other extreme where $n_s\gg n_s^\star(n_p)$.  For such
values of the parameters one is within the critical region (for small
enough $\varepsilon+\tau$), and the model exhibits strong anomalous
price fluctuations for market sizes of a few thousand agents. The
deviations from the analytical theory (which is valid only in the
limit of infinite systems) mark strong finite-size effects in the
critical region. As illustrated in the lower panel of
Fig. \ref{fig:timeseries} imposing a sufficiently large tax may in
such markets have a pronounced effect on the volatility, whereas
smaller transaction fees may influence the time-series of the market
only marginally. Fig. \ref{fig:ns60} presents a systematic account
of these effects and shows the dependence of the volatility, the
predictability and the revenue from the tax on the system size and the
tax rate $\tau$ at $n_s\gg n_s^\star(n_p)$. In particular, a significant
reduction of the market volatility can be obtained, while still keeping
the market relatively information efficient.  Furthermore, the
contribution to the tax revenue of speculators largely outweighs that
of producers and it is peaked at a value close to that where the
volatility is minimal. The effect of a tax, as shown in
Fig. \ref{fig:ns60}, also depends on the size of the market. The
volatility at low $\varepsilon+\tau$ indeed decreases with the size of
the system and approaches the theoretical line, making a tax more
effective in small than in large markets.

\begin{figure}
\begin{center}
\includegraphics[width=6cm]{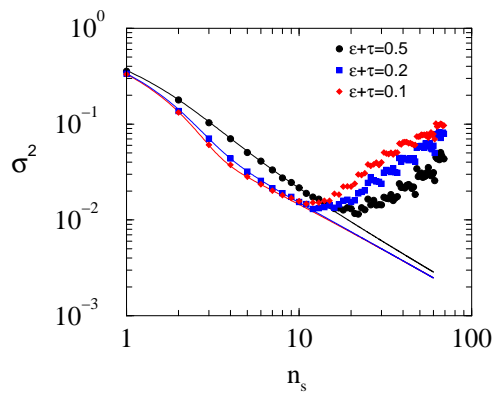}
\caption{Volatility as a function of $n_s$ at fixed $n_p=1$ for different tax rates
($\epsilon+\tau=0.1,~0.2$ and $0.5$). Markers represent data obtained from
numerical simulations with $PN_s=3000$, run for
$200P+200P/(\varepsilon+\tau)$ steps (with measurements in the second
half of this interval). An average over $3000$ samples is taken. Lines
are the corresponding predictions obtained from the analytical theory.
\label{fig:sig_vs_ns}}
\end{center}
\end{figure}

Fig. \ref{fig:sig_vs_ns} relates these two extremes and discusses the
dependence of the volatility on $n_s$ at intermediate number of
speculators for fixed $n_p=1$. We here fix the (effective) system size
by keeping $L=PN_s$ constant. For small values of $n_s$ one is then
well outside the critical region and the numerical results follow the
analytical predictions (solid lines in Fig. \ref{fig:sig_vs_ns}). As
discussed in Ref. \cite{GCMG} the simulations then deviate
systematically from the theory at large $n_s$ when the system has
entered the zone near the phase transition line. More precisely, one
finds a threshold value $\overline{n}_s(L)> n_s^\star$ so that
numerical simulations agree with the theoretical lines for
$n_s<\overline{n}_s(L)$ but deviations and anomalous fluctuations are
observed for $n_s>\overline{n}_s(L)$. As $L$ is increased
$\overline{n}_s(L)$ is found to grow as well in simulations (not shown
here), and in particular one has
$\lim_{L\to\infty}\overline{n}_s(L)=\infty$ (at $\varepsilon+\tau\neq
0$) so that the critical region vanishes in the limit of infinite
systems.

\section{Markets with evolving composition of agents}
\label{sec:changing_agents}

In the previous sections we have assumed that all traders stay in the
market for an infinite amount of of time and that their trading
strategies remain fixed forever. Individual agents have the option to
abstain from trading at intermediate times and to join the market
again at a later stage, but no agent in the setup considered so far
can actually modify his strategy vector $\{a_i^{\mu}\}$. Thus the composition
of the population of traders does not change over time. In real-market
situations however it would appear more sensible to expect some
fluctuation in the population of traders and to assume that the
market composition will evolve and/or that strategies get replaced
after some time. In the latter case, one would expect poorly
performing strategies to be removed from the market and replaced by
new ones. 

\begin{figure}
\begin{center}
\includegraphics[width=6cm]{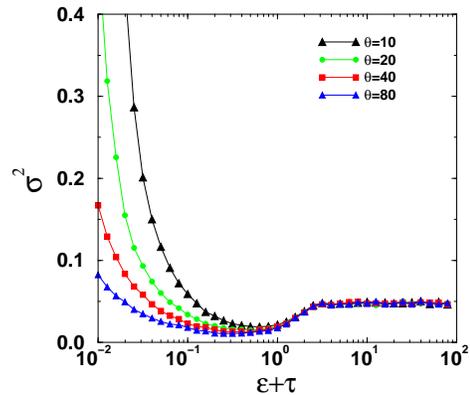}
\caption{Volatility as a function of $\epsilon+\tau$ in markets
with slowly changing composition. The parameter $\theta$ indicates the
rate at which agents are replaced; on average one replacement event
occurs in the entire population every $\theta P$ time-steps (see main
text for further details). Simulations are at $n_p=1,n_s=20$ and
$L=PN_s=5000$ with an equilibration time $T=400 P/(\epsilon+\tau)$ and
different values of $\theta$ as indicated. 
\label{fig:theta}}
\end{center}
\end{figure}

In this section we consider the simplest case of an evolving
composition of the market, namely a situation in which agents (or
equivalently their strategies) are replaced randomly, irrespective of
their performance. More precisely, at each time step each speculator
is removed with a probability $1/(\theta N_s P)$ and replaced by a new
one with randomly drawn strategy and zero initial score.  Here
$\theta$ is a constant, independent of the system size. This choice
$\theta={\cal O}(L^0)$ guarantees that the expected survival time of
any individual agent scales as $N_s P$ so that one exit/entry event
occurs in the entire population on average over a period of $\theta P$
transaction time steps. Relaxation times in Minority Games are
known to be of the order of $P$ so that the above scaling of $\theta$
results in the composition of the market changing slowly on times
comparable with those on which the system relaxes. Indeed, extensive
numerical simulations show that the behaviour of the volatility on
$\theta$ as well as that of other quantities characterizing the
collective behaviour of the system is independent of the system size
(see Fig.
\ref{fig:theta}). This is in sharp contrast with the strong finite size effects observed for $n_s \gg n_s^\star$ at a fixed composition of the population of agents (Fig. \ref{fig:ns60}). 

The main feature of the MG market with an evolving population of
agents is a pronounced minimum of the volatility as a function of
$\varepsilon+\tau$ in the crowded regime $n_s>n_s^\star$. In
particular the volatility increases as $\varepsilon+\tau$ is
decreased, even in the limit of large system sizes (in which a
corresponding system with fixed agent population would equilibrate to
the flat theoretical line as shown in Fig. \ref{fig:ns60}). This
behaviour of the system with changing agent structure can be related
to the fact that relaxation time of the GCMG scales as
$1/(\epsilon+\tau)$ \cite{Damien_private_comm}. Thus, when
$\epsilon+\tau$ is very small, the time it would take a fixed
population of agents to equilibrate collectively can be much larger
than the time scale $\theta P$ over which the market composition
changes. In this case, the market remains in a high volatility state
indefinitely because the agents do not have sufficient time to
`coordinate' and to adjust their respective behaviour as the strategy
pool represented in the evolving population of agents changes too
quickly. Fig. \ref{fig:theta} demonstrates that introducing a tax in
such markets with dynamically evolving trader structure can reduce the
volatility considerably.

\section{Conclusions}

We have shown how the theoretical picture derived for GCMG \cite{GCMG,
Book1,Book2} can be used to fully characterize the impact of a Tobin
tax on this toy model of a currency market. The main results of our study (see Fig.
\ref{fig:ns60}) is that within the GCMG the introduction of a Tobin Tax reduces the market volatility whenever the market is operating close to the critical line at $\epsilon=0$, $n_s> n_s^\star$ which marks perfect efficiency. In this region close to criticality the efficiency scales as $H\sim\varepsilon^{\alpha}$ (with a theoretically predicted exponent of $\alpha=2$) and the market is close to full efficiency $H\approx 0$. The model is known to exhibit stylized facts such as broad
non-Gaussian return distributions and volatility clustering for such
values of the model parameters. Thus the reduction of the volatility
by an additional tax is most pronounced when the market is operating
close to efficiency and exhibits anomalous fluctuations. Within the
GCMG the effect of a trading tax is due to a significant amount of
excess volatility close to a critical line. The tax $\tau$ draws the
market away from criticality in parameter space and thus reduces
volatility.

The picture is different when a tax is introduced to a market which
operates far from criticality. For a market with few speculators
($n_s<n_s^\star$) the predictability $H$ attains a finite limit when
$\varepsilon+\tau\to 0$. In this case the tax has a small effect on
volatility.

In both regimes we find that the revenue for the market maker from the
tax attains a maximal value at intermediate tax rates. In the case of
a market near criticality this occurs approximately at a tax rate at
which the reduction in volatility is largest. In both shown examples
the revenue for the market maker appears to weigh more on
institutional investors than on speculators. Measuring the revenues
from producers and speculators respectively at other values of $n_s$
and $n_p$ confirms this observation and we find $R_p>R_s$ (not shown
here). Only when the number of producers is extremely small or when
$\varepsilon<0$ can one observe instances in which the revenue from
speculators is higher than that from producers. Both an extremely
small number of producers and/or a negative value of the profit margin
$\varepsilon$ seem somewhat unrealistic in the present context so that
we do not report further details here.

Thirdly our
findings demonstrate that imposing a tax can also reduce the market
volatility in cases where the composition of the population of traders
changes slowly in time. In this case, the tax allows the agents to
reach coordinated state faster so that the market can reach a
stationary state of relatively low volatility.

Given the stylized nature of the MG, it is hard to make a connection
between the model parameter $\tau$ and an actual tax rate in a
real-world market. At any rate it seems reasonable to assume that a
realistic tax rate should be of the order or smaller than the margin
of profit of speculators, which is gauged by $\epsilon$ in the GCMG.
The optimal tax rate $\tau$ might be unrealistically large compared to
$\epsilon$. E.g. in Fig. \ref{fig:ns60} if $\epsilon=0.01$ volatility
can be substantially reduced only for tax rates $\tau$ which are more
than ten times larger.

The GCMG can at best be seen a minimalistic simplified version of a
real market. Hence our conclusions on the behaviour of the MG are at
most suggestive with respect to what might happen in the real
world. Still, the Minority Game is able to capture the interplay
between stochastic fluctuations and information efficiency in a system
of adaptive agents in a simplified way. At the same time it is
analytically solvable with the tools of statistical mechanics so that
quantitative predictions for example of the volatility can be made
based on an exact theory. In this sense, the analysis of MGs goes far
beyond the results of zero-intelligence models. Most importantly in
the present context our study provides a coherent theoretical picture
of the effects of `throwing some sand in the wheels' of markets
operating close to information efficiency.  The picture developed here
can be extended in a number of directions toward more realistic market
modes, but without giving up analytical tractability. One of the most
interesting future directions might be to endow agents with individual
wealth variables which evolve according to their relative success when
trading in the model market.

This work was supported by the European Community's Human Potential Programs
under contracts HPRN-CT-2002-00319 STIPCO and COMPLEXMARKETS. The authors
would like to thank Damien Challet and Andrea De Martino for helpful
discussions.

\section*{Appendix}
We here sketch the theoretical analysis of the model with $\tau=0$ and
general values of $\varepsilon$. The introduction of a tax $\tau$ can
be accounted for by replacing $\varepsilon\to\varepsilon+\tau$ in all
equations below. The starting point of the statistical mechanics
approach is the function
\begin{equation}
H_\varepsilon[\{\phi_i\}]=\frac{1}{P}\sum_{\mu=1}^P
  \left[\sum_{i=1}^{N}a_i^\mu\phi_i
  \right]^2+\frac{
  2\varepsilon}{P}\sum_{i=1}^{N_s}\phi_i
\label{eq:htau}
\end{equation}

\noindent of the mean activities $\{\phi_i=\avg{n_i(t)}\}$ of the
speculators $i=1,\dots,N_s$. The $\phi_i,\, i=1,\dots,N_s$ are
continuous variables within the interval $[0,1]$, recall that
producers are always active and have
$\phi_i=1,\,i=N_s+1,\dots,N=N_s+N_p$. Note that this function depends
explicitly on the strategy assignments $\{a_i^\mu\}$ so that $H_\varepsilon$
is a stochastic quantity. The strategy vectors which are fixed at the
beginning of the game correspond to what is known as `quenched
disorder' in statistical mechanics. It turns out that the learning
dynamics (\ref{eq:update}) minimizes the function $H_\varepsilon$ in terms of
the $\{\phi_i\}$ for any fixed choice of the strategy
vectors. Computing the stationary states of the model thus reduces to
identifying the minima of $H_\varepsilon$. It is here possible to
characterize these minima using the so-called replica method of
statistical physics
\cite{mpv}. A different statistical mechanics
approach is based on so-called generating functionals and deals
directly with the update dynamics (\ref{eq:update}), see
\cite{Book2}. Both methods ultimately lead to the same equations
describing the stationary states of the model, so that we here
restrict the discussion to the former approach.

In the following we give a brief sketch of the so-called replica
analysis of the model, which allows to compute the minima of the
random function $H_\varepsilon$. To this end one first introduces the
partition function
\be
Z_\varepsilon(\beta)=\int_0^1d\phi_1\cdots\int_0^1d\phi_{N_s}\exp\left(-\beta
H_\varepsilon[\phi_1,\dots,\phi_{N_s}]\right) \ee at an `annealing
temperature' $T=1/\beta$. In the limit $\beta\to\infty$ these
integrals are dominated by configurations
$\{\phi_1,\dots,\phi_{N_s}\}$ which minimize $H_\varepsilon$ so that the
evaluation of $\lim_{\beta\to\infty} Z_{\varepsilon}(\beta)$ allows one to
characterize the minima of $H_\varepsilon$.

This procedure is in general not feasible for individual realizations
of the random strategy assignments as the dependence of $H_\varepsilon$ on
the $\{a_i^\mu\}$ is quite intricate.  Instead we will compute
`typical' quantities in the limit of infinite systems, i.e averages
over the space of all strategy assignments.

The key quantity to compute here is the free energy density \be
f_\varepsilon(\beta)=-\lim_{N\to\infty}\frac{1}{\beta N}\ln Z_\varepsilon(\beta).
\ee The limit $N\to\infty$ is here taken at fixed ratios $n_s=N_s/P$ and $n_p=N_p/P$ (so that $N_s,N_p,P$ are taken to infinity as well). All relevant properties of the typical minima of $H_\varepsilon$ can be read off from the
disorder-average of $\lim_{\beta\to\infty}f_\varepsilon(\beta)$.  Using the
identity $\ln Z=\lim_{n\to 0}\frac{Z^n-1}{n}$ this problem can be
reduced to computing averages of $Z_\varepsilon^n$, corresponding to an
$n$-fold replicated systems with no interactions between the
individual copies.  This is referred to as the replica method in
statistical physics and is a standard tool for the analysis of
problems involving quenched disorder \cite{mpv,Book1,Book2}. The
averaging procedure leads to an effective interaction between the
replicas, and requires an assumption regarding the symmetry of the
solution with respect to permutations of the replicas. In principle
this symmetry may be broken, as different replica copies may end up in
different minima of $H_\varepsilon$. In the non-efficient phase of this model,instead,  the so-called `replica symmetric' ansatz is
exact, simplifying the analysis considerably. We will here not report
the detailed intermediate steps of the calculation, but will only
quote the final outcome, namely a set of closed equations for the
variables characterizing the minima of $H_\varepsilon$ (and hence the
stationary states of the GCMG). Further details of the replica
analysis are found in
\cite{Book1,Book2,GCMG}

The minima of $H_\varepsilon$ turn out to be described by two independent
variables $K$ and $\zeta$, uniquely determined from the following two
relations:
\BE
\zeta&=&\frac{1}{\sqrt{n_s(Q(\zeta,K)+n_p/n_s)}}\nonumber \\
K&=&\varepsilon\bigg[1-\frac{n_s}{2}\left(\erf[(1+K)\zeta/\sqrt{2}]-\erf(K\zeta/\sqrt{2})\right)\bigg]^{-1}
 \nonumber \\
\EE
with 
\BE
Q(\zeta,K)&=&\frac{1}{2}\ \erfc[(1+K)\zeta/\sqrt{2}]\nonumber\\
&&+\frac{1}{\zeta\sqrt{2\pi}}\left[(K-1)
  e^{-(1+K)^2\zeta^2/2}-Ke^{-K^2 \zeta^2/2}\right] \nonumber\\
&&\hspace{-3em}+\frac{1}{2}\left(K^2+\frac{1}{\zeta^2}\right)\left(\erf[(1+K)\zeta/\sqrt{2}-\erf(K\zeta/\sqrt{2})\right)\nonumber 
\EE
These equations are easily solved numerically and one obtains $K$ and
$\zeta$ as functions of the model parameters $\{n_s,n_p,\varepsilon\}$. 
The disorder-average of quantities such as the predictability $H$ or the
mean activity of the speculators
$\phi=\lim_{N\to\infty}N_s^{-1}\sum_{i=1}^{N_s}\phi_i$ can then be expressed in
terms of $K$ and $\zeta$. One finds: \BE
H&=&\epsilon^2 \frac{n_p+n_sQ(\zeta,K)}{(n_s+n_p)K^2} \nonumber \\
\phi&=&\frac{1}{2}\erfc[(1+K)\zeta /\sqrt{2}]\nonumber\\
&&+\frac{1}{\zeta\sqrt{2\pi}}\left(e^{-K^2\zeta^2/2}-e^{-\zeta^2(1+K)^2/2}
\right)\nonumber\\ &
&+\frac{K}{2}(\erf(K\zeta/\sqrt{2})-\erf[(1+K)\zeta/\sqrt{2}]) \EE.  These results are fully exact in
the thermodynamic limit, with no approximations (except for the
replica-symmetric ansatz) made at any stage. Finally, neglecting
certain dynamical correlations between agents the volatility can be
approximated as \be \sigma^2=\varepsilon^2 \frac{n_p+n_s
Q(\zeta,K)}{(n_s+n_p)K^2}+n_s\frac{\phi-Q(\zeta,K)}{n_s+n_p}. \ee As shown in the main text of
the paper this approximation is in excellent agreement with numerical
simulations (up to finite-size and equilibration effects).


\begin{thebibliography}{99}
\bibitem{nota1}
{This is an assumption on the cognitive
limitation of agents because the game itself will generate a much
richer information than just the sequence $\mu(t)$.}
\bibitem{nota2} {This
contrasts with the definition of $\mu(t)$ in the original MG, where
the information was endogenously generated by the market, with
$\mu(t)$ encoding the sign of the past $M=\log_2 P$ price
changes. Most collective properties of the model have been shown to be
affected only weakly by the origin of information. We focus here on
the simpler case of exogenous information which is analytically more
convenient. Note however that analytical approaches to MGs with
endogenous information are also feasible, but involve much more
intricate mathematics \cite{Book2}.}
\bibitem{nota3}{We here assume that each agent holds only one trading
strategy. Generalizations to multiple strategies per player are
straightforward, and can be seen not to alter the qualitative
behaviour of the model. Hence we here restrict to the simplest
case.}
\end{thebibliography}
\end{document}